\documentclass[conference]{IEEEtran}
\usepackage{cite}
\usepackage{amsmath,amssymb,amsfonts}
\usepackage{algorithmic}
\usepackage{graphicx}
\usepackage{textcomp}
\usepackage{xcolor}
\usepackage{url}
\usepackage{datetime2}
\usepackage{transparent}
\def\BibTeX{{\rm B\kern-.05em{\sc i\kern-.025em b}\kern-.08em
    T\kern-.1667em\lower.7ex\hbox{E}\kern-.125emX}}
\begin{document}

\title{Predicting Research Trends in Artificial Intelligence with Gradient Boosting Decision Trees and Time-aware Graph Neural Networks}

\author{\IEEEauthorblockN{Yichao Lu}
\IEEEauthorblockA{\textit{Layer 6 AI}\\
Toronto, Ontario, Canada \\
yichao@layer6.ai}}

\maketitle

\begin{abstract}
The Science4cast 2021 competition focuses on predicting future edges in an evolving semantic network, where each vertex represents an artificial intelligence concept, and an edge between a pair of vertices denotes that the two concepts have been investigated together in a scientific paper. In this paper, we describe our solution to this competition. We present two distinct approaches: a tree-based gradient boosting approach and a deep learning approach, and demonstrate that both approaches achieve competitive performance. Our final solution, which is based on a blend of the two approaches, achieved the $1$st place among all the participating teams. The source code for this paper is available at \url{https://github.com/YichaoLu/Science4cast2021}.
\end{abstract}

\begin{IEEEkeywords}
semantic network, link prediction, network analysis, feature engineering, graph neural network
\end{IEEEkeywords}

\section{Introduction}

A machine learning model that is able to predict the future research trends can greatly facilitate advanced research in a variety of disciplines \cite{b1,b2}. To date existing literature on predicting future research trends with machine learning remains quite limited, primarily due to the lack of a large-scale, authoritative benchmark dataset. 

The Science4cast 2021 competition \footnote{https://www.iarai.ac.at/science4cast/} aims at addressing this challenge by providing a high-quality benchmark dataset that is an order of magnitude larger than existing semantic network datasets. The dataset used in the competition consists of a growing semantic network with $8,766$ snapshots, each corresponding to a day between the beginning of 1994 and the end of 2017. Each vertex in the semantic network represents an artificial intelligence concept, and an edge is drawn between a pair of vertices when the two concepts have been investigated together in a scientific paper. Participants in this competition are asked to build a machine learning model that can predict the future edges of this evolving semantic network. 

In this paper, we present our solution to this competition. We explored two approaches: a tree-based gradient boosting approach and a deep learning approach. We demonstrate that both approaches achieve highly competitive performance. Our final solution, which is based on a blend of the two approaches, further boosts the performance, and it achieved the $1$st place among all the participating teams.

\section{Approach}

In this section, we describe our solution to the Science4cast 2021 competition. We begin by introducing the task of the competition, followed by a description of our dataset split strategy. Then we describe our feature engineering process. After that, we present two distinct approaches: a tree-based gradient boosting approach and a deep learning approach. Our final solution is based on a blend of the two approaches.  

\subsection{Task Formulation}

Let $G_{t_{1}} = (V, E_{t_{1}})$ be an undirected graph, where $V$ is the set of vertices and $E_{t_{1}} \subseteq V \times V$ is the set of observed links until time $t_{1}$. Each vertex represents an artificial intelligence concept, and edges between vertices are drawn when two concepts are investigated together in a scientific paper. Among vertex pairs that are unconnected by time $t_{1}$, we aim at predicting which edges between unconnected vertex pairs are more likely to form by time $t_{2}$. 

The following notation is used throughout the paper. Let $u, v \in V$, the binary adjacency matrix of $G$ is denoted as $A$, where $A_{u, v} = 1$ if $(u, v) \in E$ and $A_{u, v} = 0$ otherwise. For any vertex $u \in V$, let $\Gamma(u)$ be the $1$-hop neighbors of $u$, and $|\Gamma(u)|$ be the number of $1$-hop neighbors of $u$.


\begin{table*}[!t]
\caption{The list of features and their importance scores.}
\begin{center}
\begin{tabular}{lcc}
\hline
Name & Description & Importance \\ \hline
rank\_num\_neighbors\_diff\_2\_year\_u & The rank of the number of new neighbors since $2$ years prior to $t_{1}$ for vertex $u$ & 217753490.51 \\
rank\_num\_neighbors\_diff\_2\_year\_v & The rank of the number of new neighbors since $2$ years prior to $t_{1}$ for vertex $v$ & 208691110.58 \\
rank\_num\_neighbors\_diff\_1\_year\_v & The rank of the number of new neighbors since $1$ year prior to $t_{1}$ for vertex $v$ & 124923394.68 \\
rank\_num\_neighbors\_diff\_1\_year\_u & The rank of the number of new neighbors since $1$ year prior to $t_{1}$ for vertex $u$ & 108458621.54 \\
rank\_num\_neighbors\_diff\_3\_year\_u & The rank of the number of new neighbors since $3$ years prior to $t_{1}$ for vertex $u$ & 73609968.95 \\
jaccard\_index & The Jaccard index between vertices $u$ and $v$ at time $t_{1}$ & 63507171.32 \\
rank\_num\_neighbors\_diff\_3\_year\_v & The rank of the number of new neighbors since $3$ years prior to $t_{1}$ for vertex $v$ & 61741280.01 \\
jaccard\_index\_2000 & The Jaccard index between vertices $u$ and $v$ since $2020$ & 46164742.85 \\
pagerank\_score\_v & The PageRank score at time $t_{1}$ for vertex $v$ & 18960451.79 \\
pangrank\_score\_u & The PageRank score at time $t_{1}$ for vertex $u$ & 14231787.78 \\
v & The ordinal encoding for vertex $v$ & 12303301.32 \\
u & The ordinal encoding for vertex $u$ & 12156829.26 \\
jaccard\_index\_1\_year & The Jaccard index between vertices $u$ and $v$ at at $1$ year prior to $t_{1}$ & 10217562.38 \\
jaccard\_index\_diff\_3\_year & The difference between the Jaccard indices at time $t_{1}$ and at $3$ years prior to $t_{1}$ & 1164973.71 \\
jaccard\_index\_diff\_2\_year & The difference between the Jaccard indices at time $t_{1}$ and at $2$ years prior to $t_{1}$ & 784825.26 \\
pagerank\_score\_diff\_2\_year\_v &  The different between the PageRank scores at time $t_{1}$ and at $2$ years prior to $t_{1}$ for vertex $v$ & 347271.85 \\
pagerank\_score\_diff\_2\_year\_u & The different between the PageRank scores at time $t_{1}$ and at $2$ years prior to $t_{1}$ for vertex $u$ & 296226.81 \\
rank\_num\_neighbors\_v & The rank of the number of neighbor at time $t_{1}$ for vertex $v$ & 213892.33 \\
rank\_num\_neighbors\_u & The rank of the number of neighbor at time $t_{1}$ for vertex $u$ & 213258.87 \\
rank\_num\_neighbors\_2000\_v & The rank of the number of new neighbors since $2000$ for vertex $v$ & 190724.69 \\
rank\_num\_neighbors\_1\_year\_v & The rank of the number of neighbor at $1$ year prior to $t_{1}$ for vertex $v$ & 184202.64 \\
rank\_num\_neighbors\_2\_year\_v & The rank of the number of neighbor at $2$ years prior to $t_{1}$ for vertex $v$ & 182078.28 \\
rank\_num\_neighbors\_1\_year\_u & The rank of the number of neighbor at $1$ year prior to $t_{1}$ for vertex $u$ & 171340.45 \\
rank\_num\_neighbors\_3\_year\_u & The rank of the number of neighbor at $3$ years prior to $t_{1}$ for vertex $u$ & 162496.88 \\
rank\_num\_neighbors\_2\_year\_u & The rank of the number of neighbor at $2$ years prior to $t_{1}$ for vertex $u$ & 155750.99 \\
rank\_num\_neighbors\_3\_year\_v &  The rank of the number of neighbor at $3$ years prior to $t_{1}$ for vertex $v$ & 153530.45 \\
rank\_num\_neighbors\_2000\_u & The rank of the number of new neighbors since $2000$ for vertex $u$ & 143596.41 \\
jaccard\_index\_diff\_1\_year & The difference between the Jaccard indices at time $t_{1}$ and at $1$ year prior to $t_{1}$ & 120941.88 \\
pagerank\_score\_diff\_1\_year\_v & The different between the PageRank scores at time $t_{1}$ and at $1$ year prior to $t_{1}$ for vertex $v$ & 103631.39 \\
pagerank\_score\_diff\_1\_year\_u & The different between the PageRank scores at time $t_{1}$ and at $1$ year prior to $t_{1}$ for vertex $u$ & 100832.93 \\
jaccard\_index\_3\_year & Jaccard index between vertices $u$ and $v$ at $3$ years priors to $t_{1}$ & 98789.44 \\
pagerank\_score\_diff\_3\_year\_u & The different between the PageRank scores at time $t_{1}$ and at $3$ years prior to $t_{1}$ for vertex $u$ & 86917.48 \\
pagerank\_score\_diff\_3\_year\_v & The different between the PageRank scores at time $t_{1}$ and at $3$ years prior to $t_{1}$ for vertex $v$ & 83766.08 \\
pangrank\_score\_2\_year\_u & The PageRank score for vertex $u$ at $2$ years prior to $t_{1}$ & 43588.77 \\
pagerank\_score\_1\_year\_v & The PageRank score for vertex $v$ at $1$ year prior to $t_{1}$ & 42649.46 \\
pangrank\_score\_1\_year\_u & The PageRank score for vertex $u$ at $1$ year prior to $t_{1}$ & 39638.87 \\
jaccard\_index\_2\_year & The Jaccard index between vertices $u$ and $v$ at $2$ years priors to $t_{1}$ & 26012.81 \\
pagerank\_score\_2\_year\_v & The PageRank score for vertex $v$ at $2$ years prior to $t_{1}$ & 11689.22 \\
pagerank\_score\_3\_year\_v & The PageRank score for vertex $v$ at $3$ years prior to $t_{1}$ & 9604.56 \\
pangrank\_scores\_3\_year\_u & The PageRank score for vertex $u$ at $3$ years prior to $t_{1}$ & 7482.16 \\
jaccard\_index\_diff\_2000 & The difference between the Jaccard index at time $t_{1}$ and the Jaccard index since $2000$ & 2.85 \\
\hline

\hline
\end{tabular}
\label{table:feature_engineering}
\end{center}
\end{table*}
\subsection{Dataset Split}
In the competition test set, $t_{1}$ is set to \DTMdisplaydate{2017}{12}{31}{} and $t_{2}$ is set to \DTMdisplaydate{2020}{12}{31}{}. Equivalently, the task asks for the prediction of which scientific pairs of concepts will be investigated by scientists over $3$ years. Our dataset split strategy is to create the training set and the validation set that mimic the test set. This ensures a high correlation of performance of the model on the validation set and the test set. Specifically, for the training set we set $t_{1}$ to \DTMdisplaydate{2011}{12}{31}{} and set $t_{2}$ to \DTMdisplaydate{2014}{12}{31}{}, while for the validation set we set $t_{1}$ to \DTMdisplaydate{2014}{12}{31}{} and set $t_{2}$ to \DTMdisplaydate{2017}{12}{31}{}.

For the training set and the validation set, we sample $N$ vertex pairs that are unconnected by time $t_{1}$ and the labels are determined by whether edges between unconnected pairs will form by time $t_{2}$. We set $N$ to be $10,000,000$ both for the training set and the validation set. While a larger training set typically leads to better performance, there exists a trade-off between performance and efficiency since a larger training set also incurs heavier computational cost. We find setting $N$ to $10,000,000$ results in a good balance between performance and efficiency.

The competition dataset is highly imbalanced with only $0.178\%$ newly connected edges ($3,724,986$ new edges out of $2,092,398,869$ unconnected vertex pairs) formed between $2014$ and $2017$. To deal with the scarcity of positive samples, we include all positive samples when constructing the training set and the validation set. This results in $1,275,503$ positive samples in the training set, and $3,724,986$ positive samples in the validation set. The negative samples are then uniformly sampled from the remaining unconnected vertex pairs. We experimented with alternative sampling strategies for negative samples, e.g., sampling unconnected vertex pairs according to vertex degrees to include more ``hard negatives'' in the dataset \cite{b3}, but did not see gains as expected. We also experimented with a data augmentation approach, which exploits the order invariance of the task, i.e., the model should output the same link prediction score for the vertex pair $(u, v)$ and the vertex pair $(v, u)$. Briefly speaking, we double the size of the training set by swapping the order of the vertices in each pair. We found that this data augmentation approach only brings about marginal gains in terms of performance at the cost of much longer training time.

\begin{figure*}[t]
  \centering
  \includegraphics[width=\textwidth]{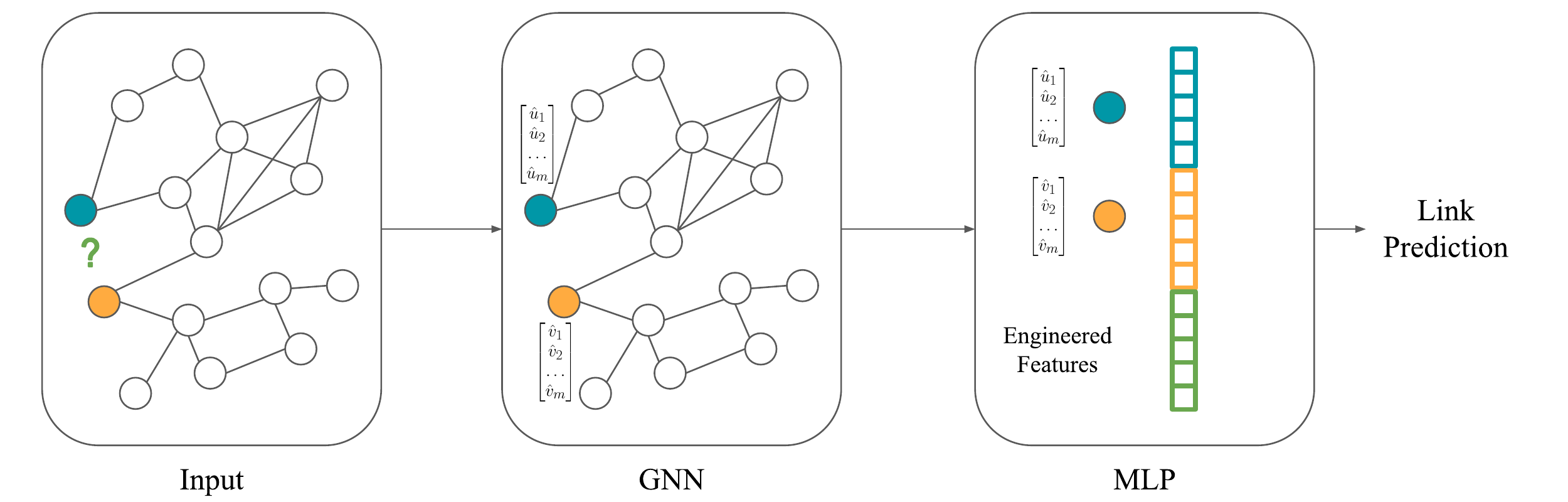}
  \caption{High-level Architecture of the Time-aware Graph Neural Network.}
  \label{figure:diagram}
\end{figure*}
\subsection{Feature Engineering}
Feature engineering plays an important role in the winning entries of various machine learning competitions \cite{b4,b4_1,b4_2}. While deep neural networks have become the de facto approach to representation learning in domains such as computer vision \cite{b5,b5_1,b5_2} and natural language processing \cite{b6,b6_1}, machine learning models still rely heavily on feature engineering with domain expertise when it comes to dealing with tabular data \cite{b7}.

We conducted extensive feature engineering with the aim of capturing the centralities of the vertices, the proximity between vertex pairs, and their evolution over time. The final set of features that we used can be partitioned into two groups: vertex features and pairwise features, which we describe in what follows.

Vertex features are derived using the number of neighbors and the PageRank score \cite{b9} of the vertex, which reflect the vertex centrality. In order to model the evolution of the vertex centrality over time, we additionally compute the centrality of the vertex (in terms of the number of neighbors and the PageRank score) at $1$, $2$, and $3$ years prior to $t_{1}$. The change of centrality of the vertex is captured by taking the difference between the centrality measures at time $t_{1}$ and that at $1$, $2$, and $3$ years prior to $t_{1}$. We also compute the centrality measures on a subset of the full semantic network, where edges that were formed prior to $2000$ are discarded, which encourages the centrality measures to be more focused on more recent edges. For vertex features derived using the number of neighbors of the vertex, we apply the rank transformation, which replaces the features with their ranks, or average ranks in case of ties. This is because the number of neighbors of the vertex is constantly increasing over time, and we need a normalization procedure to adjust these features so that the features within different dataset splits are comparable. Rank transformation works well because it is a monotonic transformation that preserves the order among the features and the distribution of the rank transformed features remains unchanged over time. Vertex features derived from the PageRank score of the vertex do not require the rank transformation since PageRank scores are already normalized.

Pairwise features are derived using the Jaccard index \cite{b10}, which is defined as the number of common neighbors between $u$ and $v$ divided by the total number of neighbors of $u$ and $v$: 

\begin{equation}
    J(u, v) = \left\{ 
        \begin{aligned}
            & \frac{|\Gamma(u) \cap \Gamma(v)|}{|\Gamma(u) \cup \Gamma(v)|}, & |\Gamma(u) \cup \Gamma(v)| > 0 \\
            & \text{NaN}, & |\Gamma(u) \cup \Gamma(v)| = 0
        \end{aligned}
        \right.,
\end{equation}

where we use NaN to indicate missing values. The number of common neighbors between $u$ and $v$ can be obtained by computing the square of the adjacency matrix:

\begin{equation}
    |\Gamma(u) \cap \Gamma(v)| = (A^{2})_{u, v}.
\end{equation}

A larger Jaccard index indicates a higher proximity between vertices, and thus a higher chance of forming edges between them. Similar to vertex features, we model the evolution of the proximity between vertex pairs by computing the Jaccard indices at $1$, $2$, and $3$ years prior to $t_{1}$ and their differences to the Jaccard index at time $t_{1}$.

The list of all features that we used in this competition is presented in Table \ref{table:feature_engineering}. We additionally show the importance score for each feature and sort the features accordingly. The importance score for each feature is obtained from a trained Light Gradient Boosting Machine (LightGBM) \cite{b11} by computing the total gains of splits that use the feature. We can see that the features corresponding to the number of new neighbors of the vertices $1$, $2$, and $3$ years prior to $t_{1}$ are the most important ones. This is because these features are very strong indicators of the current research trends. In addition, the Jaccard index is also a very effective feature since it measures the proximity between the vertex pairs. We can also see that the PageRank score is more useful than the number of neighbors of the vertex as it is a better metric of vertex centrality.

\subsection{A Gradient Boosting Approach}
In our tree-based gradient boosting approach, we use the Light Gradient Boosting Machine (LightGBM), which is a highly efficient implementation of the gradient boosting decision tree (GBDT) algorithm. We experimented with other GBDT implementations, such as XGBoost \cite{b12} and CatBoost \cite{b13}, and found that LightGBM achieves the best performance. 

We train LightGBM to minimize the binary cross entropy loss between its predictions and the ground truths. LightGBM has also implemented the LambdaMART algorithm \cite{b14}, which is a state-of-the-art Learning-to-Rank (LTR) technique \cite{b15}. Training with LambdaMART results in better validation AUC than training with the binary objective but does not work well in the blending stage. This is because LTR models output uncalibrated scores. Therefore we did not use it in our final solution. In order to combat overfitting due to the scarcity of positive samples, we set \textit{num\_leaves} to 16 and set \textit{max\_depth} to 4 to limit the growth of the trees. We also set \textit{bagging\_fraction} to 0.8 and set \textit{feature\_fraction} to 0.9, which means that we randomly select $90\%$ of the data and $80\%$ of the features before training each tree. We train LightGBM for $10,000$ iterations with a learning rate of $0.01$, and stop training when the validation AUC does not improve in the last $100$ iterations. 

Since decision trees are invariant to monotonic transformations of features and can handle missing values without imputation, we directly feed raw features into LightGBM. Note that we directly feed the ordinal encoding of the vertices $u$ and $v$ into LightGBM, despite the fact that they are essentially categorical features. A common practice to deal with categorical features in GBDT algorithms is to one-hot encode the features. However, given the high cardinality of the vertices, we found that ordinal encoding results in the best performance.



\subsection{A Deep Learning Approach}

In our deep learning approach, we employ a time-aware graph neural network to learn vertex representations on dynamic semantic networks. The learned vertex representations for the vertex pair are then concatenated with the engineered features and fed into a multi-layer perceptron (MLP) which outputs the final link prediction score; see Figure \ref{figure:diagram}. 

We propose to apply an exponential decay on the edge weights of the semantic network so as to encourage the graph neural network to pay more attention to more recently formed edges. This also improves the time-awareness of the graph neural network since it can now recover the exact time that a particular edge is formed from the edge weight. The adjacency matrix with exponentially decayed weights $\hat{A}$ is defined as 

\begin{equation}
    \hat{A}_{u, v} = \left\{ 
        \begin{aligned}
            & e ^ {-0.0001 (t_{1} - t_{u, v})}, & (u, v) \in E \\
            & 0, & (u, v) \notin E
        \end{aligned}
        \right.,
\end{equation}

where $t_{u, v}$ denotes the most recent time prior to $t_{1}$ that an edge between $u$ and $v$ was formed, and $t_{1} - t_{u, v}$ denotes the number of days between $t_{u, v}$ and $t_{1}$.

Our graph neural network is composed of $3$ graph convolution blocks, where each block takes the output of the previous block as its input except for the first block. We concatenate the engineered vertex features with a trainable vertex embedding matrix as the input to the first block of the graph neural network. Each graph convolution block consists of a SAGEConv layer \cite{b16}, a Batch Normalization layer \cite{b17}, a ReLU activation layer, and a skip connection \cite{resnet}:

\begin{equation}
    h^{l} = \text{ReLU}(\text{BatchNorm}(\text{SAGEConv}(h^{l - 1}, \hat{A}))) + h^{l - 1},
\end{equation}

\begin{table}[t]
\caption{Validation performance for LightGBM, the time-aware graph neural network, and the blend.}
\begin{center}
\begin{tabular}{l|c|c}
\ Approach & AUC & Training time \\ \hline \hline
\ LightGBM \ &  \ \ \ 0.90293041 \ \ \ & \textbf{5 minutes} \\ \hline
\ GNN & 0.90297814 & 97 minutes \\ \hline
\ Blend & \textbf{0.90314956} & 102 minutes \\ \hline
\end{tabular}
\label{table:validation}
\end{center}
\end{table}

where $h^{l}$ denotes the vertex representations learned by the $l$-th graph convolution block.

After obtaining the learned vertex representations from the output of the last graph convolution block, we index the representations corresponding to vertices $u$ and $v$ and concatenate them with the engineered pairwise features. The concatenated features are then fed into a $5$-layer MLP with densely connected layers \cite{densenet}. We use ReLU activation units for intermediate layers of the MLP, and use the sigmoid activation unit in the final layer, which squashes the output value to the interval of $(0, 1)$ to represent the probability of existence of the new edge.

Deep neural networks are very sensitive to the scale of the inputs, and hence the need for standardization. We preprocess the engineered features before feeding them into the neural network by first filling all the missing values with $0$ and then standardizing each feature through dividing the mean subtracted values by the standard deviation. The training objective for the neural network is the same as that for LightGBM, which is to minimize the binary cross entropy loss. The model is optimized with the Adam optimizer \cite{b18} with batch size $512$, learning rate of $1e-4$, and a weight decay of $1e-3$. We apply a dropout of $0.2$ to prevent the model from overfitting and train the model for $30$ epochs.

\subsection{Blending}

Our final solution is based on a blend of the gradient boosting approach and the deep learning approach. We use blending instead of other ensembling techniques such as stacking since we observe that our models have very high variances and relatively low biases. For blending, we begin by taking the third power of the link prediction scores given by each model so as to amplify the relative difference \cite{b19} and then compute the weighted average of the transformed scores as the final link prediction score for each vertex pair.

\section{Experiments}

\subsection{Dataset and Evaluation Metric}

The dataset used in the competition is a dynamic semantic network characterizing the content of the scientific literature in the field of artificial intelligence between 1994 and 2017. The network consists of $64,719$ vertices, each representing an artificial intelligence concept. Edges between vertices are drawn when two concepts are investigated together in a scientific paper, and the timestamp of the edge records the date of publication of the scientific paper.




The evaluation metric used in the competition is AUC (area under the curve–receiving operating characteristic), which is a widely adopted metric to evaluate the performance of a binary classifier on different confidence thresholds. ROC is a probability curve that visualizes change of the true positive rate with respect to the false positive rate on different threshold values. The AUC represents probability that a positive sample is ranked higher than a negative sample \cite{b20, auc}. An AUC equal to 1 is ideal and represents that the model can perfectly distinguish positive samples from negative samples.

\subsection{Results and Analysis}

\begin{table}[t]
\caption{Ablation results for the graph neural network.}
\begin{center}
\begin{tabular}{l|c|c}
Approach & \# SAGEConv & AUC \\ \hline \hline
MLP & - & 0.89503802 \\ \hline
MLP w/ vertex embedding & - & 0.89712115 \\ \hline
GNN w/ binary adjacency matrix & 3 & 0.90203951 \\ \hline
GNN w/ exponential decay & 1 & 0.90217735 \\ \hline
GNN w/ exponential decay & 2 & 0.90254369 \\ \hline
GNN w/ exponential decay & 3 & \textbf{0.90297814} \\ \hline
GNN w/ exponential decay & 4 & 0.90278343 \\ \hline
GNN w/ exponential decay & 5 & 0.90273941 \\ \hline
\end{tabular}
\label{table:ablation}
\end{center}
\end{table}

The validation performance of LightGBM, the time-aware graph neural network, and the blend is presented in Table \ref{table:validation}. We can see that the graph neural network is able to achieve slightly better performance, while LightGBM is significantly more efficient. The blend is able to further boost the performance, albeit by a small margin. 

We additionally performed an ablation study on the graph neural network; see Figure \ref{table:ablation}. We observe that directly applying the simple MLP on top of our engineered features fails to achieve comparable performance with LightGBM and the graph neural network. Equipping the MLP with a vertex embedding module improves its performance since it can now learn vertex representations in a data-driven manner. The vanilla graph neural network, which performs graph convolution operations over the binary adjacency matrix, has a significant performance advantage over the MLP since it is able to learn vertex representations aggregated within a certain neighborhood region. Our exponential decayed adjacency matrix further improves the performance by enabling the model to learn time-aware vertex representations. We also studied the impact of the number of graph convolution blocks on the performance of the graph neural network. Empirically we found that the graph neural network achieves the best performance when the number of graph convolution blocks is set to $3$. This may indicate that the neighborhood information within 3-hop is most useful in the semantic network of this competition.


The final leaderboard result \footnote{https://www.iarai.ac.at/science4cast/competitions/2021-core/?leaderboard} for the top $5$ teams is presented in Table \ref{table:leaderboard}. Our solution achieved the $1$st place among all the participating teams, with more than $1.3\%$ of relative improvement over the $2$nd place team.

\section{Discussion and Future Work}

In this section we describe a few possible avenues of future improvements to our solution.

When creating the training set, we fix $t_{1}$ to \DTMdisplaydate{2011}{12}{31}{} and $t_{2}$ to \DTMdisplaydate{2014}{12}{31}{}. This limits the number of positive samples in the training set and makes the model prone to suffer from overfitting. A possible way to obtain more positive samples in the training set is to vary $t_{1}$ and $t_{2}$. We can obtain a number of training sets with different $t_{1}$ and $t_{2}$, and combine them into a larger training set with more positive samples. This can potentially lead to further improvement to our solution since the performance of machine learning models is largely dependent on the size of the training dataset.

In addition, we demonstrate that using a weighted adjacency matrix can lead to better performance of the graph neural network. In our solution, the weights for the edges are defined using a handcrafted formula. An alternative approach is to learn the edge weights with deep neural networks. For example, we can employ the graph attention network (GAT) \cite{gat} to dynamically learn the edge weights through an attention mechanism. Given sufficient amounts of training data, we typically find handcrafted features to be outperformed by the representations learned with deep neural networks.

Furthermore, in our graph neural network approach, the embedding vectors for the vertices are fixed through time. This limits the capability of the vertex embedding module to capture the evolution of the vertex over time. 
Alternatively, we can try incorporating a time dependent embedding module into our graph neural network \cite{tgn}. 

\begin{table}[t]
\caption{The final leaderboard result for the top $5$ teams.}
\begin{center}
\begin{tabular}{c|c|c}
Rank & Team & AUC \\ \hline \hline
\textbf{1} & \textbf{oahciy} & \textbf{0.9283886} \\ \hline
2 & Hash Brown & 0.9273865 \\ \hline
3 & SanatisFinests2 & 0.9221236 \\ \hline
4 & Bacalhink & 0.9185331 \\ \hline
5 & nimasanjabi & 0.9184558 \\ \hline
\end{tabular}
\label{table:leaderboard}
\end{center}
\end{table}

\section{Conclusion}

We present our solution to the Science4cast 20201 competition, which is based on a blend of a tree-based gradient boosting approach and a deep learning approach. Our approach yields highly competitive results, achieving the $1$st place out of all the participating teams.

\end{document}